\documentclass[aps,prfluids,preprint,superscriptaddress,tightenlines]{revtex4-2}

\usepackage{amsmath,amssymb}
\usepackage{graphicx}
\usepackage{xcolor}
\usepackage{booktabs}

\usepackage{stackengine}
\usepackage{ulem}

\definecolor{mygreen}{rgb}{0.,.5,0.}

\begin{document}


\title{Numerical study of Lagrangian velocity structure functions using
acceleration statistics and a spatial-temporal perspective}

\author{Rohini Uma-Vaideswaran}
\affiliation{
School of Aerospace Engineering, \\ Georgia Institute of Technology, Atlanta, GA 30332, USA}
\email{rvaideswaran3@gatech.edu}
\author{P. K. Yeung}%
\affiliation{
 Schools of Aerospace Engineering and Mechanical Engineering, Georgia Institute of Technology, Atlanta, GA 30332, USA
}%


\vspace{1.5cm}

\date{\today}
\def\ktime {{\tau_\eta}}

\vspace{1.5cm}

\begin{abstract}

\ 

A fundamental relation in Lagrangian Kolmogorov theory is concerned with 
inertial range scaling of the second-order velocity structure function over intermediate time lags
at sufficiently high Reynolds numbers. Significant  theoretical
support for asymptotic constancy of the scaling constant~($C_0$) 
is known, but limitations in the range of time scales
accessible in direct numerical simulation make unambiguous
testing of the scaling challenging. In this paper, direct numerical simulations of forced isotropic turbulence at 
Taylor-scale Reynolds numbers  
between 140 and 1300 are used to improve
understanding in this subject, including scaling
at time lags shorter and longer than those 
associated with inertial-range like  behavior. Uncertainties arising from modest simulation time spans in the high Reynolds number data
are addressed by expressing the velocity structure function
in terms of the acceleration autocorrelation, which suggests that $C_0$ 
may be sensitive to effects of Lagrangian intermittency but does
not rule out asymptotic constancy at Reynolds numbers beyond
those that may be feasible in simulations in the foreseeable future. The Lagrangian velocity increment is examined further from a spatial-temporal perspective, as a combination of convective (spatial) and local (temporal) contributions, 
which are subject to a strong but incomplete
mutual cancellation dependent on Reynolds number and time lag.
The convective contribution is strongly influenced by the 
particle displacement, which is driven by large-scale dynamics
and can thus grow into inertial range dimensions in space
within just a few Kolmogorov time scales, without
fully satisfying classical Lagrangian inertial-range requirements.
An overall conclusion in this work is that both the limited 
range of time scales (narrower than that for length scales) and 
the effects of particle displacements have significant roles 
in the observed behavior of the second-order Lagrangian velocity structure function.
\end{abstract}

\maketitle

\newpage

\section{Introduction}

\def\be { \begin{equation} }
\def\ee { \end{equation} }
\def\uu {\mathbf{u}}
\def\ur {\mathbf{r}}
\def\ux {\mathbf{x}}
\def\ua {\mathbf{a}}
\def\ktime {{\tau_\eta}}
\def\re {{ R_\lambda }}
\def\uv {\mathbf{v}}
\def\epsav {{\langle\epsilon\rangle}}

\baselineskip=.2in

It is well known that important and complementary 
insights about turbulence can be obtained from
both the Eulerian viewpoint of a fixed observer and the
Lagrangian viewpoint of an observer moving with the instantaneous flow.
In particular, two fundamental quantities are the
Eulerian and Lagrangian velocity increments, i.e.,
\begin{eqnarray}
	\Delta_r\uu & = & \uu(\ux+\ur)-\uu(\ux), \\
	\Delta_\tau\uu^+ & = & \uu^+(t+\tau)-\uu^+(t),
\end{eqnarray}
where $\ur$ is a separation vector, $\tau$ is a
time lag, the superscript $+$ denotes quantities
following particle trajectories, and explicit
dependence on time $t$ in Eq.~1 may be omitted if the
turbulence is statistically stationary.
Statistics of $\Delta_r\uu$ are crucial to the
classical Kolmogorov similarity hypotheses~\cite{K41}(K41 for short) and refinements therefrom~\citep{K62,frisch95,SA97}, while
the statistics of
$\Delta_\tau\uu^+$ are crucial in many attempts to
extend K41 theory~\citep{MY.II}
and/or to develop
intermittency corrections~\citep{biferale.2008,yu2010lagrangian} for
Lagrangian quantities where applicable.
The properties of $\Delta_\tau\uu^+$
are also crucial for stochastic modeling, where a key task is to
predict the statistics of $\uu^+(t+\tau)$ when given $\uu^+(t)$,
with modeling of the particle acceleration $\ua^+(t)$
being crucial for capturing
Reynolds number dependence~\citep{Sawford.1991}.

Due to challenges in both experiments and simulations,
despite the importance of the subject,
the statistics of $\Delta_\tau\uu^+$ are less well characterized 
or understood than those of $\Delta_r \uu$. A prime
example is the
second-order Lagrangian structure function
$D_2^L(\tau)=\langle(\Delta_\tau \uu^+)^2\rangle$,
where angled brackets denote averaging over a large particle
population, and stationarity in time implies
$D_2^L(\tau)$ is a function of $\tau$ only.
For isotropic (including locally isotropic) turbulence,
direct use of K41 theory predicts
\be
\tfrac{1}{3}\langle(\Delta_\tau \uu^+)^2\rangle = C_0\epsav \tau,
\ \ \ (\ktime \ll \tau \ll T_L),
\label{eq:C0}
\ee
where
$\epsav$ is the mean energy dissipation rate,
$\ktime$ is the Kolmogorov time scale, $T_L$
is the Lagrangian integral time scale of the particle velocity,
and $C_0$ is called the Lagrangian Kolmogorov constant, which
is supposedly universal at sufficiently high Reynolds numbers.
However, in practice, plots of 
$\tfrac{1}{3}\langle(\Delta_\tau \uu^+)^2\rangle/(\epsav \tau)$
typically (e.g., in ~\citep{yeung2002,SY.2011})
do not exhibit a distinct plateau. Instead,
most data sources show a smooth local peak,
with some sensitivity to Reynolds number,
in the range of 5--10 $\ktime$. As the Reynolds
number and hence the time-scale ratio $T_L/\ktime$ increases,
this range apparently satisfies the criterion
$\ktime\ll\tau$ less clearly than it does for  $\tau\ll T_L$
~\citep{YPS06}.
Lagrangian quantities also tend to show stronger intermittency, which is clear from many studies~\citep{VSB.1998,YPKL07,bentkamp19natcom}
showing the fluid particle acceleration (proportional to $\Delta_\tau \uu^+$ when $\tau$ is small) having a stronger propensity for extreme values than known for Eulerian velocity gradients in space.

In this paper, our goal is to advance understanding of 
the observed behavior of the Lagrangian
second-order structure function in (forced) stationary isotropic 
turbulence, by employing two
complementary approaches. The first approach is to consider
the velocity increment as the integral of the fluid particle
acceleration, which gives a double integral (for each coordinate component)
\be
\langle(\Delta_\tau u^+)^2\rangle
= \int_0^\tau \int_0^\tau
\langle a^+(t+t')a^+(t+t'')\rangle dt'dt'',
\ee
where, by stationarity,  the result is independent of 
the reference time $t$.
This expression is similar in spirit to a formula that
gives mean-square displacement in terms of the velocity
autocorrelation, dating back to Taylor~\citep{GITaylor}.
The form of the integral implies that both the variance and 
autocorrelation of the acceleration have important
roles in the statistics of the Lagrangian velocity
increment, potentially over a substantial period of time.

The second approach is to 
consider $\Delta_\tau \uu^+$ as a spatial-temporal increment, recognizing that $\uu^+(t+\tau)$ differs from its prior value $\uu^+(t)$ for two reasons, namely that, during the time interval $\tau$, (a) the instantaneous Eulerian velocity field has
evolved in time, while (b) the particle has moved to a new location $\ux^+(t+\tau)$. In particular, we may write
\begin{align}\label{eq:stdecomp}
    \Delta_\tau \uu^+(t) = [\uu(\ux^+(t+\tau),t+\tau) - \uu(\ux^+(t),t+\tau)] + [\uu(\ux^+(t),t+\tau) - \uu(\ux^+(t),t)] \ ,
\end{align}
where the first square bracket on the r.h.s is a spatial increment over a length scale corresponding to the particle displacement $\ell(\tau) = |\ux^+(t+\tau)-\ux^+(t)|$, and the second is a purely temporal increment taken at the particle's prior position $\ux^+(t)$. We refer to these as convective and local increments, respectively.
The former is more complex than the Eulerian spatial increment $\Delta_r\uu$, because the particle displacement $\ell(\tau)$ is a random variable whose values can spread over a wide range of scales. 
Since advective transport is driven by the large scales, even in a short time ($\mathcal{O}(\ktime)$), fluid particles can move by a distance much larger than
$\eta$, causing the convective increment to quickly pick up samples with spatial separations in the inertial range or even beyond. The particle displacement, together with the random-sweeping hypothesis
by Tennekes \cite{Tennekes}, is thus an important parameter in the underlying physics.

In this paper, we present results from
direct numerical simulations (DNS)
of forced isotropic turbulence.
The Taylor-scale Reynolds number
range covered is from 140 to 1300, which may be considered
to provide \citep{ISY.2017}
reasonable representation of inertial range behavior for Eulerian
statistics, although a higher Reynolds number
is necessary for the corresponding Lagrangian
scaling~\citep{SY.2011}.
It will be seen that, due to computational
resources being finite, the time span of some of our higher-resolution
simulations are quite modest --- but we will show that the key
conclusions in this paper are not significantly compromised by this limitation.
An analysis of the statistics of the Lagrangian velocity increment
in terms of the fluid particle acceleration
provides some understanding why 
the scaling suggested in Eq.~\eqref{eq:C0} is unlikely
to hold perfectly even at high Reynolds numbers.
The apparently strong yet incomplete cancellation
between convective and local contributions to
the velocity increment is studied using
conditional sampling, which provides a physical
mechanism for
the emergence of strong intermittency in 
$\Delta_\tau\uu^+$. In addition, the statistics of distance
traveled by a fluid particle over a time period $\tau$
also helps explain why any plateau resembling Eq.~\eqref{eq:C0}
is likely to be quickly truncated.

It may be noted that decompositions similar to that 
in Eq.~\ref{eq:stdecomp} have been employed
previously in the context of bridging Lagrangian
and Eulerian statistics, e.g. in 
\cite{kamps2009exact,homann2009bridging,kamps2009lagrangesche}
which used formal relations between Eulerian and Lagrangian PDFs
in terms of transition probabilities.
The authors of \cite{lalescu2018tracer}
also showed using single-particle dispersion and random sweeping arguments
that Eulerian increments with separations distributed over a wide
range play an important role. Numerical experiments on particles moving through 
frozen velocity fields with no local acceleration
\cite{chevillard2005intermittency,friedrich2009statistics}
also showed that a coupling between local and convective
contributions is necessary to reproduce the strong
Lagrangian intermittency observed in DNS. 
However, Lagrangian intermittency is a subject where
high Reynolds number is especially important, in fact more so
than for its Eulerian counterpart. In this paper, results on the Reynolds number dependence of the spatial-temporal properties conditioned on tracer particle displacements cover a broad range of Taylor-scale Reynolds numbers
extending up to three times larger than in some of the prior literature.

The remaining sections are organized as follows. In Sec.~II, we provide information on the numerical approach and the Lagrangian simulation database. 
Sec.~III contains results on the velocity structure function and
an analysis addressing the role of the acceleration autocorrelation.
In Sec.~IV, we focus on a spatial-temporal perspective and the role 
of particle displacements on Lagrangian intermittency. Conclusions are summarized in Sec.~V.

\section{Computational Approach and Simulation Overview}

We have performed DNS using well-known Fourier pseudo-spectral methods for 3D forced isotropic turbulence on an $N^3$ periodic domain~\cite{rogallo}. 
The forcing scheme
used is that of~\cite{DY2010} where the energy spectrum
(but not individual Fourier modes) at the lowest few wavenumber shells
is frozen in time. The fluid particle velocity
is obtained by cubic-spline interpolation~\citep{YP.1988} from the
Eulerian velocity field according to the relation
$\uu^+(t)=\uu(\ux^+(t),t)$, while particle positions
are advanced in time using a second-order Runge Kutta scheme.
Lagrangian time series are written out at intervals of $\ktime/20$
or smaller for subsequent post-processing.
A large particle count $N_P$ is important in
simulations at higher Reynolds numbers and grid resolutions
for better sampling of extreme events (e.g., in the fluid acceleration).

The largest simulations reported here at resolution $12288^3$ used
2048 nodes on the supercomputer {\it Frontera} at the Texas Advanced
Computing Center (TACC). 
To ensure  high code performance,
particle-related communication costs are minimized
by distributing the particles dynamically
among multiple parallel processes based on their instantaneous
positions~\citep{buaria2017highly}. The most
expensive interpolation-related operation
is the generation of cubic spline coefficients
from the Eulerian velocity field, 
which involves solving tridiagonal
systems of simultaneous equations with periodic boundary conditions.
In departure from the remote memory-access approach
in~\citep{buaria2017highly}, 
for greater inter-platform portability, we make spline coefficients near 
the boundaries of adjacent sub-domains available to each
parallel process using ghost layers and one-sided communication. These features have resulted in overall costs depending little on the particle count, with $N_p$ of order $10^8$ or $10^9$ achieved economically.

The spatial-temporal decomposition in Eq.~\eqref{eq:stdecomp} 
requires knowledge of quantities of the type
$\uu(\ux^+(t_0),t_0+\tau)$, which is the particle velocity at the
end of a time interval $[t_0,t_0+\tau]$ but calculated at the
particle position at time $t_0$. This type of information
can be difficult to obtain unless
particle positions at prior time instants are retained,
which would lead to a heavy burden on
computer memory requirements.
However, stationarity implies the statistics of
$\uu(\ux^+(t_0),t_0+\tau)$ depend only on $\tau$, and not $t_0$.
This allows us to simply set $t_0=0$ and store the initial
particle positions, while continuing to perform interpolation
at those initial positions at later times as
the simulation proceeds. Since this approach 
does not require any additional spline coefficients
to be computed, any extra cost is minimal.

\def\re {{ R_\lambda }}
\def\kme {{ k_{max}\eta }}
\def\bb  {{ \ \ \ \ }}
\setlength{\tabcolsep}{8pt}
\begin{table}[h]  
\small
{  
\begin{tabular}{rrrrrrrr}
\midrule
    \bb $N^3 $  & \bb $N_P$  & \bb $R_{\lambda}$ & \bb $\langle\epsilon\rangle$ & \bb $T/\tau_{\eta}$ & \bb $T/T_L$ & \bb $T_L/\tau_{\eta}$ & \bb $\kme$  \\\midrule
\addstackgap{$256^3$}  & 2~M & 140 & 1.43 & 162.63 & 12.0 & 13.50 & 1.34\\\hline
\addstackgap{$1536^3$} & 54~M & 390 & 1.23 & 181.93  & 5.70  & 31.93 & 2.07 \\
\hline
\addstackgap{$3072^3$} & 54~M & 650 & 1.44 & 18.16  & 0.33 & 54.49 & 1.99 \\
\hline
\addstackgap{$6144^3$} & 192~M & 1000 & 1.46 & 18.27  & 0.21 & 87.01 & 1.99 \\\hline
\addstackgap{$12288^3$} & 192~M & 1300 & 1.37 & 15.27 & 0.14 & 111.02 & 2.69 \\
\midrule
\end{tabular}
}
\caption{DNS parameters: grid resolution, particle count (in multiples of $1~{\rm M}=1024^2$), Taylor-scale Reynolds number, mean dissipation rate, simulation time span $T$ in units of $\tau_\eta$ and $T_L$, the time scale ratio $T_L/\ktime$, and the non-dimensional resolution parameter $\kme$.}
\label{table:DNS}
\end{table}

Table~\ref{table:DNS} provides a brief summary
of the parameters of the present simulations. As in~\citep{YPS06}, the Reynolds
number is increased by reducing the viscosity
with forcing parameters (that control
the large scales) unchanged. This leads to the mean energy dissipation rate ($\epsav$) 
being largely insensitive to Reynolds number,
although, the variations that do occur may be
significant enough to introduce ambiguities when $\epsav$ is used
as a normalizing factor (as in Eq.~\eqref{eq:C0}). The forcing scheme we used is that of
\cite{DY2010}, which was specifically motivated by a desire
to reduce temporal varibility of $\epsav$,
with instantaneous values of $\epsav$ differing only modestly from
long-time averages recorded in long previous simulations that used
the stochastic forcing scheme of \cite{EP88}.


We have used mostly values of $N$ that include a single factor of 3,
which is conducive for optimal code performance 
given the hardware characteristics of {\it Frontera}.
Except for the lowest Reynolds number, we have maintained
small-scale resolution at the level of 
$\kme$ (where $k_{max}=\sqrt{2}N/3$ is the highest resolvable wavenumber on an $N^3$ grid after dealiasing) at least 2 (which gives a grid spacing $\Delta x$ very close to 1.5 $\eta$). 

The main limitation in our higher Reynolds number simulations is their relatively short duration.
Ideally, $T$ should be large compared to $T_L$.
Unfortunately,
although computer memory and speed have
been advancing very fast,
the cost of long simulations at the
grid resolutions needed at high Reynolds numbers
is increasing even faster.
Studies of small-scale physics in the Eulerian frame
can conceivably be conducted using multiple short simulations
with independent initial conditions~\cite{YR.2020,Yetal2025}.
However, for Lagrangian calculations 
there are other considerations.
First, since $T_L$ is determined by the forcing parameters, 
which are held fixed, we can use the same
value of $T_L$ for all simulations, including shorter ones where
the integral defining $T_L$ is not well-converged.
Second, with inertial range scaling
being our initial motivation, $T$ should at least extend beyond
the range where the normalized structure function displays
its peak value (between 5--10 $\ktime$, per~\citep{SY.2011}),
with the effects of finite $T$ to be examined carefully. As seen in Table 1, our simulations meet this requirement.

\def\re {{ R_\lambda }}
\def\kme {{ k_{max}\eta }}
\def\bb  {{ \ \ \ \ }}
\def\avdiss {{\langle\epsilon\rangle}}
\def\ktime {{\tau_\eta}}

\section{Velocity structure function and the role of acceleration autocorrelation}

\def\re {{ R_\lambda }}
\def\kme {{ k_{max}\eta }}
\def\bb  {{ \ \ \ \ }}
\def\avdiss {{\langle\epsilon\rangle}}
\def\ktime {{\tau_\eta}}

As stated earlier, our main goal is to explain
the observed numerical behavior of the second-order Lagrangian
velocity structure function, including the value of $C_0$ in
Eq.~\eqref{eq:C0}. Theoretical studies based on
expectations concerning long-time behavior
\cite{BS.1991}, 
Lagrangian multifractal theory \cite{Borgas.1993},
as well as stochastic modeling \cite{SY.2011}
have suggested that $C_0$ approaches
a constant at infinitely high Reynolds number.
In principle,
it is possible to estimate such an asymptotic
value using extrapolation formulas with coefficients
based on DNS data at the (necessarily finite) Reynolds
numbers available.
Analyses given in  \cite{BS.1991} also indicate
the long-time behavior of the acceleration
autocovariance, and hence potentially
the Kolmogorov-scaled acceleration variance, i.e.,
\be
a_0=\tfrac{1}{3}\langle \ua\cdot\ua\rangle/(\epsav^{3/2}\nu^{-1/2}),
\label{eq:azero}
\ee
(where $\ua$ is the acceleration vector and $\nu$ is the kinematic viscosity)
has a significant role  in the inertial range.
We note that, as computations have advanced over the years,
acceleration is now known to deviate from classical Kolmogorov scaling,
with $a_0$ increasing systematically
with the Reynolds number~\citep{VSB.1998,VY.1999} as a result
of intermittency. 

\begin{figure}[h]
    \centering    \includegraphics[width=0.9\linewidth]{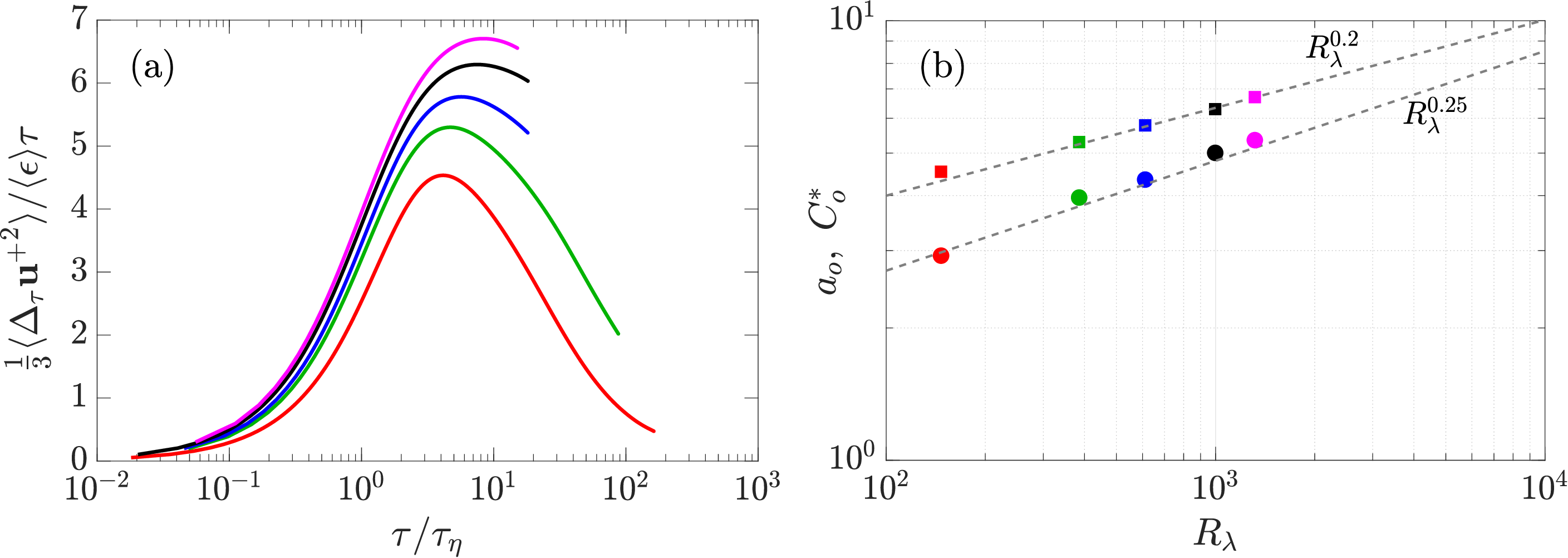}
    \caption{\small (a) Normalized second-order Lagrangian structure function 
    at $\re$ from 140 to 1300 per Table I (red, green, blue, black, magenta respectively) (b) peak of the curve in (a), denoted by $C_0^*$ (squares) and  $a_0$ (circles). Dashed gray lines indicate power laws fits for comparison: $\re^{0.2}$ (for $C_0^*$) and $\re^{0.25}$ (for $a_0$).}
    \label{fig:strfn}
\end{figure}

In practice, a reasonable way to estimate $C_0$ is to plot the second-order
Lagrangian function, normalized by $\epsav \tau$, and look for the peak value,
here denoted by $C_0^*$, at intermediate time lags.
Figure~\ref{fig:strfn}(a) shows such a plot,
which reaches a smooth peak at a $\tau/\ktime$ in the range
roughly 5--8. The peak appears to be shifted slightly towards
higher $\tau/\ktime$ at higher $\re$, but 
there is no clearly recognizable plateau.
In frame (b), we show both the height of this peak, 
as well as $a_0$, versus the Reynolds number.
The data on $a_0$ appear to follow closely a power law of slope 0.25,
which is consistent with recent work in the literature
\cite{buaria2023lagrangian}.

The situation concerning
$C_0^*$ is less clear. 
If a power-law is assumed then the data in 
Fig.~\ref{fig:strfn} suggests that
a shallow power law of slope 0.2, smaller than
the slope of 0.25 for $a_0$, may provide a reasonable fit.
However, it would not be surprising if
an equally-good fit can be obtained with
an interpolation formula that suggests an eventual
leveling-out at a value higher than 7 
suggested in \cite{SY.2011} --- perhaps
in the neighborhood of 8 --- at Reynolds numbers well beyond the reach of DNS in the foreseeable future.
For consistency, we have also verified that, in Fig.~\ref{fig:strfn},
comparison with data points
on $C_0^*$ from long simulations in the past at $\re\sim$ 140 to 1000
in \cite{SY.2011} does not lead to different conclusions.


As noted earlier,
the limited time span of some of the higher Reynolds number
simulations is a source of statistical uncertainty for both
the velocity structure function and its associated
autocorrelation function
$\rho_L(\tau)$.
In particular, the calculation of $\rho_L(\tau)$ 
for time lags $\tau$ comparable to $T$ may involve
many temporal increments which are overlapping 
and therefore lacking in  statistical independence, while
the ambiguity between so-called biased
and unbiased estimators in time-series analysis~\citep{Priestley}
can also be appreciable.
However, a useful check can be
made using the acceleration autocorrelation, which
has shorter time scales and hence can be obtained more
reliably without $T$ being comparable to $T_L$. 
In particular, in the same manner as the mean-square displacement
being expressible in terms of the velocity autocorrelation,
the mean-square of the velocity increment can be expressed
in terms of the acceleration autocorrelation as (see also \cite{BS.1991,pope.book})
\begin{eqnarray}
\langle (\Delta_\tau u^+)^2\rangle
& = & 2  \langle a^2\rangle \int_0^\tau (\tau-s)~\rho_a(s)~ds \\
& = & 2 \langle a^2\rangle
\left[\tau \int_0^\tau \rho_a(s)~ds
-\int_0^\tau s~\rho_a(s)~ds
\right],
\label{eq:C0_1}
\end{eqnarray}
which follows from assuming stationarity and performing
an integration by parts. 
We refer to the two integrals in Eq.~\eqref{eq:C0_1}
as $I_0$ and $I_1$ respectively.
Both are controlled
by the shape of the function $\rho_a(s)$, which
has a remarkably
robust zero-crossing at 
$s\approx 2.2\ktime$~\cite{YP.1989,YPKL07}, followed by a negative loop which ends in
a slow asymptotic approach to zero.
In principle, the negative loop should be consistent
with the acceleration having a zero
integral time scale, which is a consequence of
the acceleration being the time derivative of
a stationary process (the velocity). However,
if $T$ is not sufficiently long, the integral 
$\int_0^T\rho_a(s)~ds$ will be positive.
At small $\tau$, both integrals increase from zero
until the zero-crossing of $\rho_a(\tau)$. Subsequently
they decrease, with $I_0$ approaching zero while remaining positive,
but in contrast $I_1$ becomes negative at a small $\tau$, 
say $\tau_1$, and eventually approaches a negative constant,
consistent with the requirement that 
$\langle (\Delta_\tau u^+)^2\rangle$ approaches $2\langle u^2\rangle$
as $\tau\rightarrow\infty$.

From Eq.~\eqref{eq:C0_1}, multiplying by $1/\tau$ and differentiating with respect to $\tau$, results in
\be
\frac{d}{d\tau}
\left[
\frac{\langle (\Delta_\tau u^+)^2\rangle}{\tau}
\right]=
2\langle a^2\rangle
\frac{1}{\tau^2}\int_0^\tau s~\rho_a(s)~ds,
\label{eq:C0_2}
\ee
which indicates the peak of the normalized
structure function occurs at $\tau=\tau_1$, where $I_1$ becomes negative. The peak value itself is given by
\be
\langle (\Delta_\tau u^+)^2\rangle/\tau
= 2 \langle a^2\rangle \int_0^{\tau_1} \rho_a(s)~ds.
\label{eq:C0_3}
\ee
It seems reasonable to suppose the integral
$\int_0^{\tau_1} \rho_a(s)~ds$ to be mainly determined by
information within the first few $\ktime$, and potentially scale with $\ktime$ itself.
If we write
\be
\int_0^{\tau_1} \rho_a(s)~ds = \gamma~\ktime,
\label{eq:gammadef}
\ee
then normalizing both sides
of Eq.~\eqref{eq:C0_3} by the Kolmogorov variables gives
\be
\tilde{C}_0^* = 2~a_0~\gamma,
\label{eq:C0_gamma}
\ee
where $\tilde{C}_0^*$ represents $C_0^*$ recovered from $\rho_a(\tau)$ and $\langle a^2\rangle$  for comparison with a direct
calculation from the normalized second-order structure 
function. Since we are expressing velocity increments
as integrals of the acceleration, it is not surprising that,
due to memory effects, $a_0$ may have a long-lasting
effect on velocity increment statistics and hence
the value of $\tilde{C}_0^*$.
We have seen earlier in Fig.~\ref{fig:strfn}(b)
that, within the data range available, 
$C_0^*$ increases with Reynolds number, but does so at
a slower rate than $a_0$. This suggests $\gamma$
decreases with $\re$. 

\begin{figure}[h]
\centerline{
\includegraphics[width=0.9\textwidth]{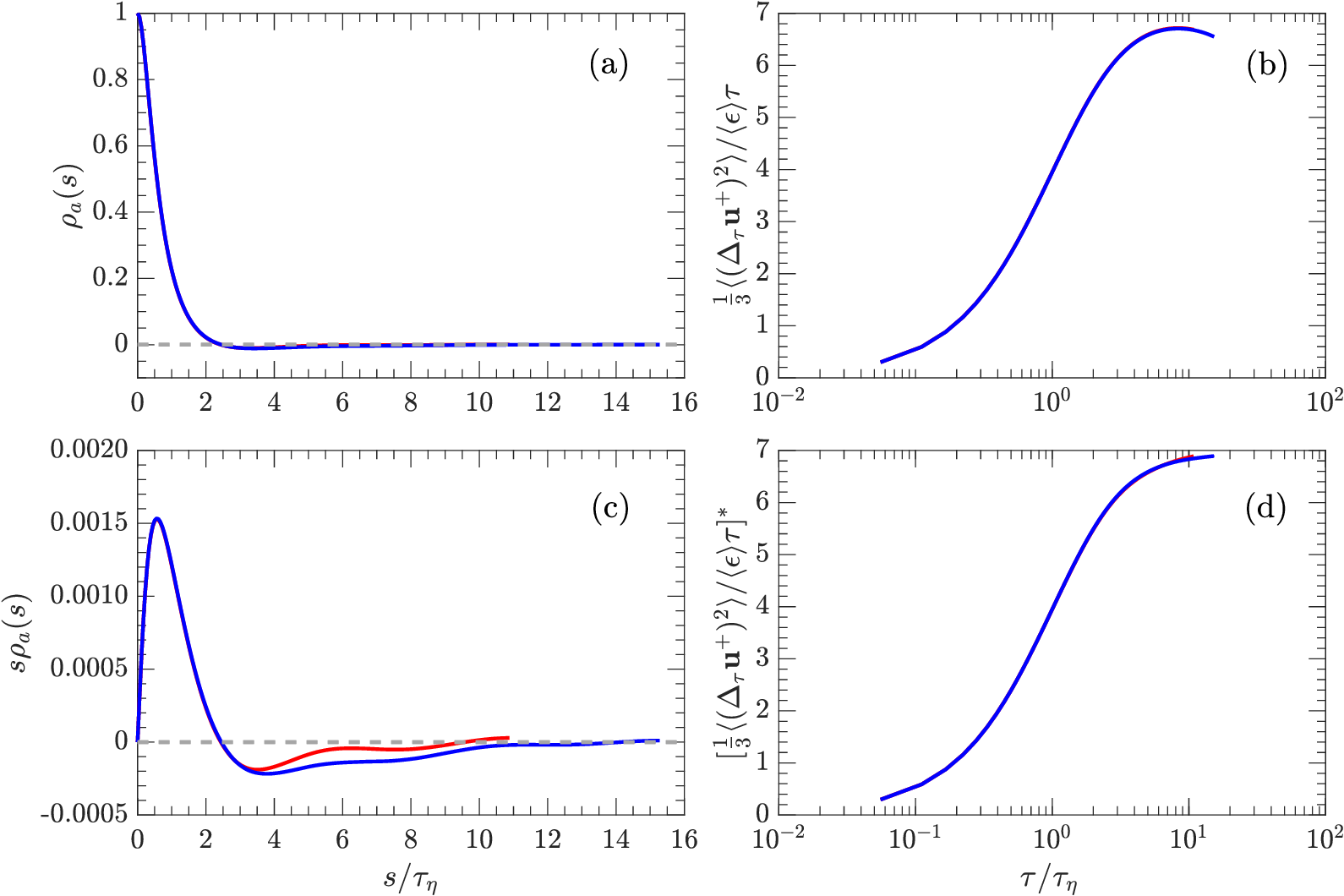}
}
\caption{\small Results from $12288^3$ simulation at the highest $\re$ 
($\sim$1300)  (a) acceleration autocorrelation and  (c) same as (a) 
but multiplied by time lag; structure function from (b) a direct calculation 
or (d) recovered from the quantities $\rho_a(\tau)$ and $\langle a^2\rangle$. Lines in red and blue are for data processed up to 
	about ${11}~\ktime$ 
	and   ${15}~\ktime$ respectively. Except in portions of (c), they nearly coincide.}
\label{fig:re1300}
\end{figure}

It should be noted that accurate calculation of the parameter  $\gamma$
according to Eq.~\eqref{eq:gammadef} can be challenging if a simulation
is limited in length, which can affect the numerical values of  $\rho_a(s)$
at larger values of $s$, and the estimation of $\tau_1$ which sets
the upper integration limit. In the two longest (and lowest $\re$) simulations
(see Table~I) of this work the values of $C_0^*$ and $\tilde{C}_0^*$
differ by less than 1\%. However, in the other three shorter but higher  $\re$
simulations the difference is more noticeable, being 7\% in the  $\re\sim$ 1300 data.
In fact, if a simulation is not long enough for the integral $I_1$ on the r.h.s 
of Eq.~\eqref{eq:C0_2} to become negative, accurate estimation of the time lag $\tau_1$ 
where the peak of the normalized structure function occurs would be difficult.

To assess more closely the impact of the simulation time span on the evaluation of the normalized structure function and its peak,
in Fig.~\ref{fig:re1300} we use data at $\re\sim$ 1300 to compare
statistics obtained by processing only the first 11 $\ktime$ versus the full 15 $\ktime$. In frame~(a), 
the acceleration autocorrelation appears 
quite robust
--- but multiplication by $\tau$ (in frame~(c)) reveals
the presence of significant noise occurring between
approximately 4-8 $\ktime$.
This noise, which is more evident if $T$ is short, makes the accurate assessment of $\gamma$ in 
Eq.~\eqref{eq:C0_gamma} challenging in short simulations.
Nevertheless, it is worth noting that the
difference apparent between the red and blue lines
in frame (b) does not translate into a significant
discrepancy between corresponding lines in frame (d).
On the other hand, comparison between frames (b) and (d)
suggests that in our $\re\sim$ 1300 data,
$\tilde{C}_0^*$ may be slightly higher than $C_0^*$,
reflecting a limitation associated with short simulation
time spans as already discussed.

\def\re {{ R_\lambda }}
\def\kme {{ k_{max}\eta }}
\def\bb  {{ \ \ \ \ }}
\def\ellav {{ \langle\ell\rangle }}

\section{A spatial-temporal perspective}

Although our main focus is on finite-time increments
per Eq.~\eqref{eq:C0}, 
it is clear that at small $\tau$,
Eq.~\eqref{eq:stdecomp} gives $\Delta_\tau 
\uu^+\approx (\ua_C+\ua_L)\tau$, where $\ua_C \equiv (\uu\cdot\nabla)\uu$ and $\ua_L \equiv \partial\uu/\partial t$ are, 
respectively, the convective and local contributions to the 
material derivative of the velocity in the Navier-Stokes equations. A strong mutual cancellation between $\ua_C$ and $\ua_L$ is 
known~\citep{YP.1989,TVY.2001} in accordance with the so-called  ``random-sweeping'' hypothesis by Tennekes~\cite{Tennekes}, which assumes that small-scale turbulence structure is 
passively advected by the large-scale motions. 
There is recent
work in the literature \cite{buaria2023lagrangian} focused on 
the Reynolds number scaling of the statistics of $\ua_C$ and $\ua_L$ computed as single-time quantities in the Eulerian frame. 
In addition, the authors of
~\cite{lalescu2018tracer} used arguments 
based on single particle dispersion and random-sweeping, 
to show that Lagrangian velocity increment 
statistics arise from the mixing of Eulerian increments 
over a wide range of scales through an effective Lagrangian dispersion. Here, we consider more generally two-time statistics in the Lagrangian frame.

\subsection{Random-sweeping and mutual cancellation of the velocity increments}

We denote the convective and local velocity increments defined in Eq.~\eqref{eq:stdecomp} by $\uv_C(\tau)$ and $\uv_L(\tau)$, respectively, such that 
\begin{align}\label{eq:stdecomp2}
\Delta_\tau \uu^+(\tau) = \uv_C(\tau) + \uv_L(\tau). \
\end{align}
At small $\tau$,
$\uv_C(\tau)$ and $\uv_L(\tau)$ are
subject to the same strong (but incomplete) mutual cancellation
as noted for the convective and local accelerations above.
Clearly, the rate of growth of $\Delta_\tau \uu^+$ with time
lag $\tau$ is sensitive to how
the degree of mutual cancellation changes, which in turn
is governed by the nature of the joint statistical
distribution of the vectors $\uv_L$ and $\uv_C$,
including their degree of geometric alignment.

\begin{figure}[h]
    \centering
    \includegraphics[width=\textwidth]{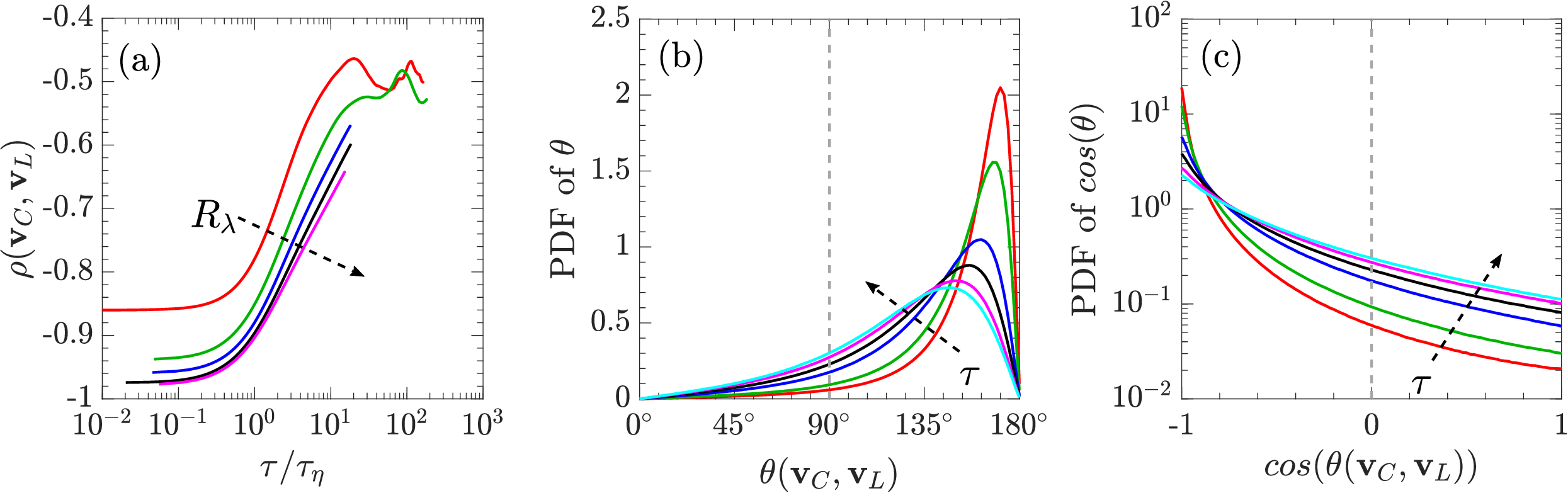}
    \caption{\small (a) Evolution of $\rho(\uv_L,\uv_C)$ in time  at $\re\sim140, 390, 650, 1000, 1300$ (increasing in the direction of the arrow), (b) and (c) PDFs of $\theta$, and $\cos(\theta)$ respectively, at $\re\sim390$ and time lags $\tau/\ktime\approx 1, 2, 4, 8, 16, 32$, (in the order red, green, blue, black, magenta, cyan).}
    \label{fig:corcof}
\end{figure}

Figure~\ref{fig:corcof} shows (a) how the correlation coefficient $\rho(\uv_C,\uv_L)$ evolves with Kolmogorov-scaled time lag at different Reynolds numbers, and (b, c) how the PDFs of the alignment angle $\theta$ between $\uv_C$ and $\uv_L$ (and its cosine) at a given Reynolds number evolves for $\tau$  ranging from smaller than $\ktime$ to comparable to $T_L$. 
Initially, $\rho(\uv_C,\uv_L)$ is close to $-1.0$, with the alignment angle having a high probability of being close to $180^\circ$, which is especially true at higher $\re$. 
Since random-sweeping depends on a disparity of scales,
it is expected to become weaker as the time lag increases
and the velocity increments begin 
to reach beyond the dissipative scales.
Indeed, the degree of anti-correlation begins to weaken noticeably
from $\tau\approx 0.5 \ktime$ onwards. However, it should be noted, as seen
in curves representing the longer simulations in Fig.~\ref{fig:corcof}, that
the asymptotic value of this autocorrelation
is not zero, but close to~-0.5. This asymptotic level can be explained by noting that,
at large $\tau$, while 
$\uu(\ux^+(t+\tau),\tau)$, $\uu(\ux^+(t),t)$  
and $\uu(\ux^+(t),t+\tau)$ are all independent and
identically distributed (with mean variance  $\langle u^2\rangle$),
$\uv_L$ and $\uv_C$  are not independent since both still
contain $\uu(\ux^+(t),t+\tau)$, with different signs. Algebraically, the covariance $\langle\uv_C\cdot\uv_L\rangle$ approaches $-\langle u^2\rangle$ while the variances of both $\uv_C$ and $\uv_L$ approach $2\langle u^2\rangle$. We have also verified that, in our longest simulation, at
$\re\sim$ 390, the PDFs of both the alignment angle and its cosine
evaluated at $\tau/\ktime$ over 180 differs only minimally from
the last time step ($\tau/\ktime\approx 32$) shown in both 
frames (b) and (c).

Although the degree of mutual cancellation between $\uv_L$ and $\uv_C$ 
at small $\tau$ is strong, complete cancellation is not possible since 
$\uv_C$ contains irrotational contributions 
but $\uv_L$ does not, and also because 
complete cancellation would result in a zero total acceleration.
In addition,
the observations from Fig.~\ref{fig:corcof} are subject to the caveats
that (because of non-Gaussianity) correlation coefficients do not
provide a full picture of how $\uv_L$ and $\uv_C$ depend on each
other statistically, and that statistics on alignment angles
do not distinguish between events where
the magnitudes of $\uv_L$ and $\uv_C$ may be large or small.

\begin{figure}[h]
\centerline{\ \ \ 
\includegraphics[trim={0 0 0 0},clip,width=0.92\textwidth]{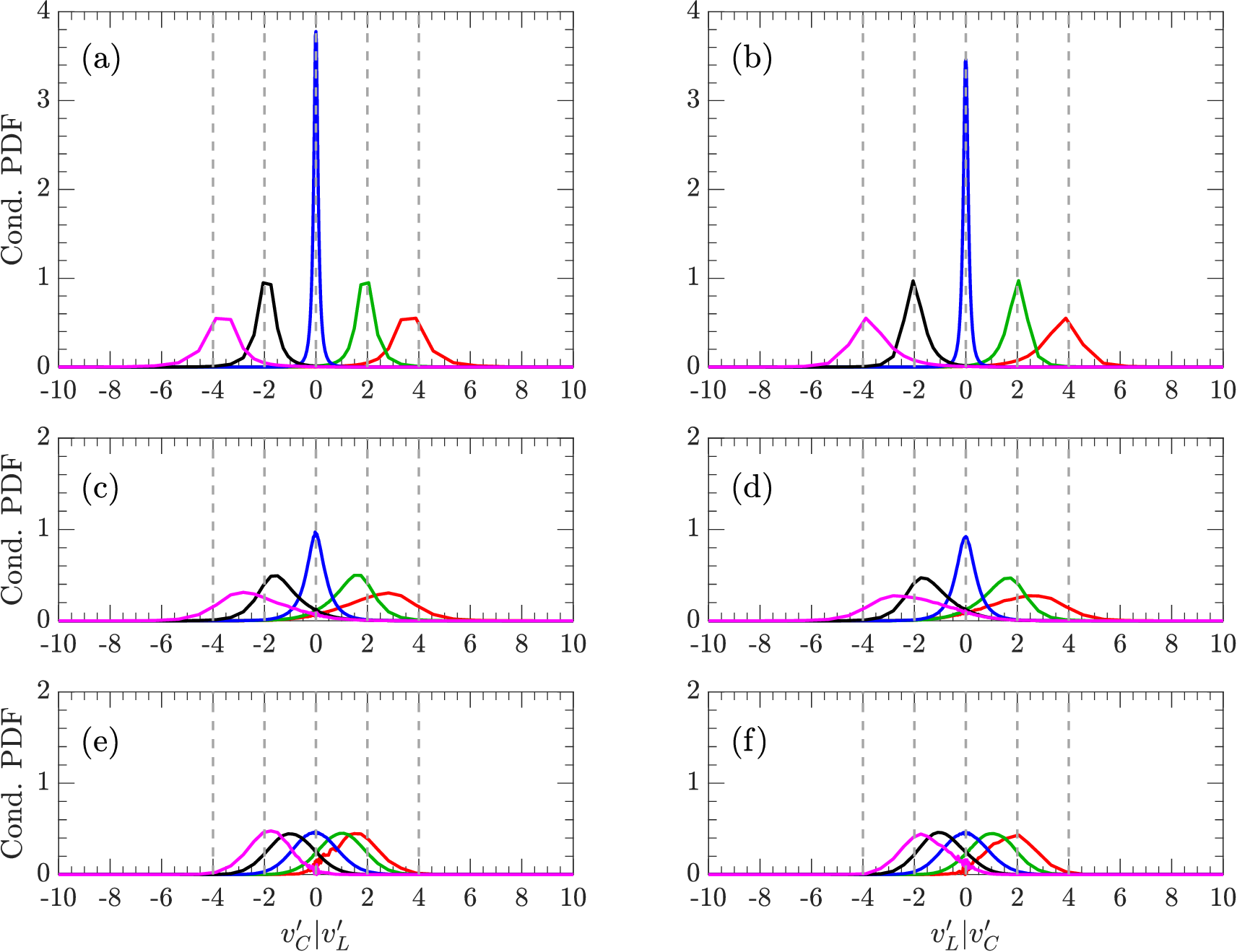}
}
\caption{\small
Conditional PDFs of $v_C^{'}$ given $v_L^{'}$ (left column) and
$v_L^{'}$ given $v_C^{'}$ (right column), with
primes denoting  normalization by unconditional root-mean-square (RMS.),
using data at $\re\sim$ 390 on a $1536^3$ grid.
Each frame consists of 5 lines, with the conditioning variable
at -4, -2, 0, 2, 4 times the RMS values (red, green, blue, black, 
magenta) from the mean. Top, middle, and bottom rows show results 
at $\tau/\ktime \approx 0.1$, $4$ and $100$, respectively. 
}
\label{fig:cpdf_re390}
\end{figure}

\begin{figure}[h]
\centering
\includegraphics[trim={0 0 0 0},clip,width=0.92\textwidth]{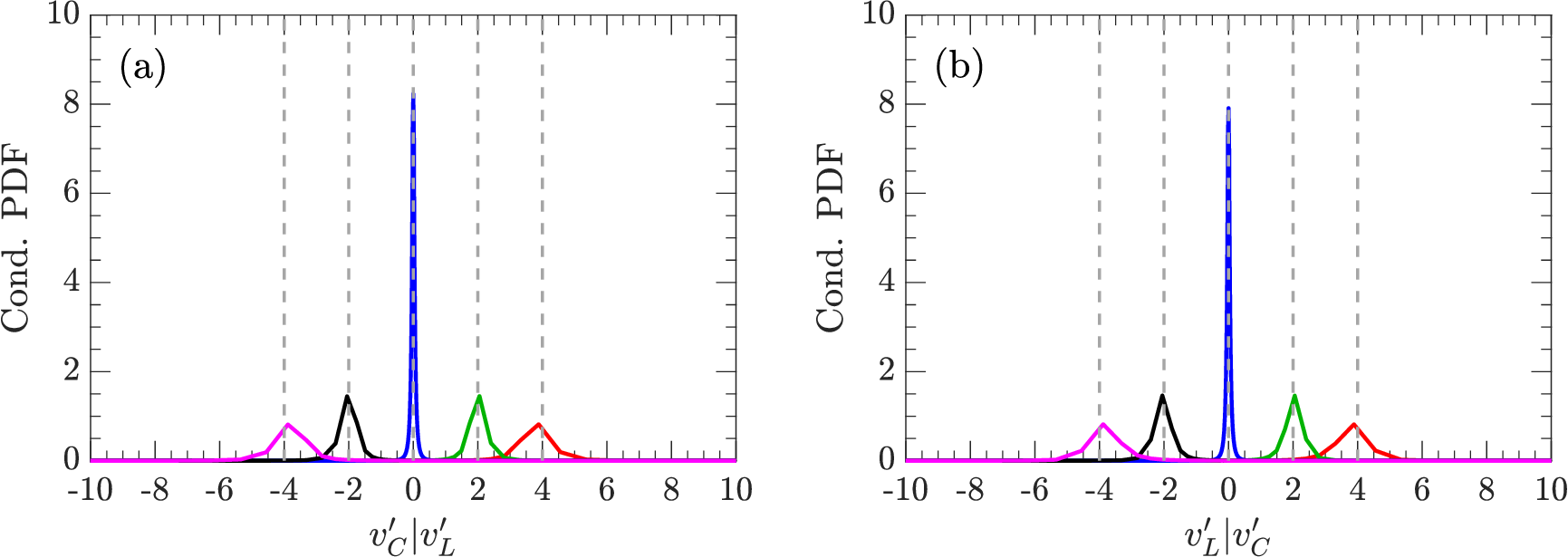}
\caption{\small Same as top row Fig.~\ref{fig:cpdf_re390}, but for data at $\re\sim1300$, on a $12288^3$ grid.
}
\label{fig:cpdf_re1300}
\end{figure}

For a deeper understanding of the joint statistics
of $\uv_L$ and $\uv_C$, we can examine
the conditional PDFs of each variable given the other,
at different values of $\tau/\ktime$.
Figure~\ref{fig:cpdf_re390} shows representative results
at  $\re\sim$ 390. 
Although conditional PDFs can be very noisy,
here, we need only focus on fluctuations 
on the order of several RMS values.
If mutual cancellation is complete,
then the PDF of $\uv_C$ given $\uv_L$
should show a sharp spike at $\uv_C = -\uv_L$,
or $v_C'=-(\sigma_C/\sigma_L)v_L'$
when normalization by the respective unconditional
RMS values (denoted by $\sigma_C$ and $\sigma_L$)
is taken into account.
However, such sharp spikes are not well observed.
Even at small $\tau$, although large values of the conditioning variables lead to the other variable taking mostly values
of the opposite sign, the conditional PDFs
are spread out over a finite range instead of
a narrow spike. The magenta line of
frame (b) shows a slight bias towards smaller values of $v_C^{'}$
when conditioned on $v_L^{'}$,
suggesting that
$\uv_L$ and $\uv_C$ do not have the same statistics.
As $\tau$ increases and the anti-correlation
between $\uv_L$ and $\uv_C$ weakens, the spikes in blue
gradually fade. 
However, close observation of data at 
larger values of the conditioning variable
(lines in red and magenta in frames (e) and
(f)) show samples do exist
where the two increments are of the same sign.
Comparison between lines in red or magenta 
in frames (a,c,e) or (b,d,f)
also show that, as $\tau$ increases, 
any preferential association between large $\uv_L$ and
large $\uv_C$ is gradually lost.
The conditional PDFs of $\uv_C$ given $\uv_L$
and of $\uv_L$ and $\uv_C$ appear to differ
appreciably only at very small $\tau$, where
intermittency is the strongest.

Figure~\ref{fig:cpdf_re1300} shows results similar to the top 
row of Fig.~\ref{fig:cpdf_re390}, but at a substantially higher
$\re$~(1300).
The most noticeable contrast is that
the spike at early times given near-zero values
of the conditioning variable has become taller as well as 
narrower. This enhanced spike is consistent with
prior findings that the random-sweeping hypothesis
holds better at higher $\re$.
The bias noted above in Fig.~\ref{fig:cpdf_re390}(b)
is also apparently weaker as $\re$ increases.

\begin{figure}[h]
    \centering
    \includegraphics[width=\textwidth]{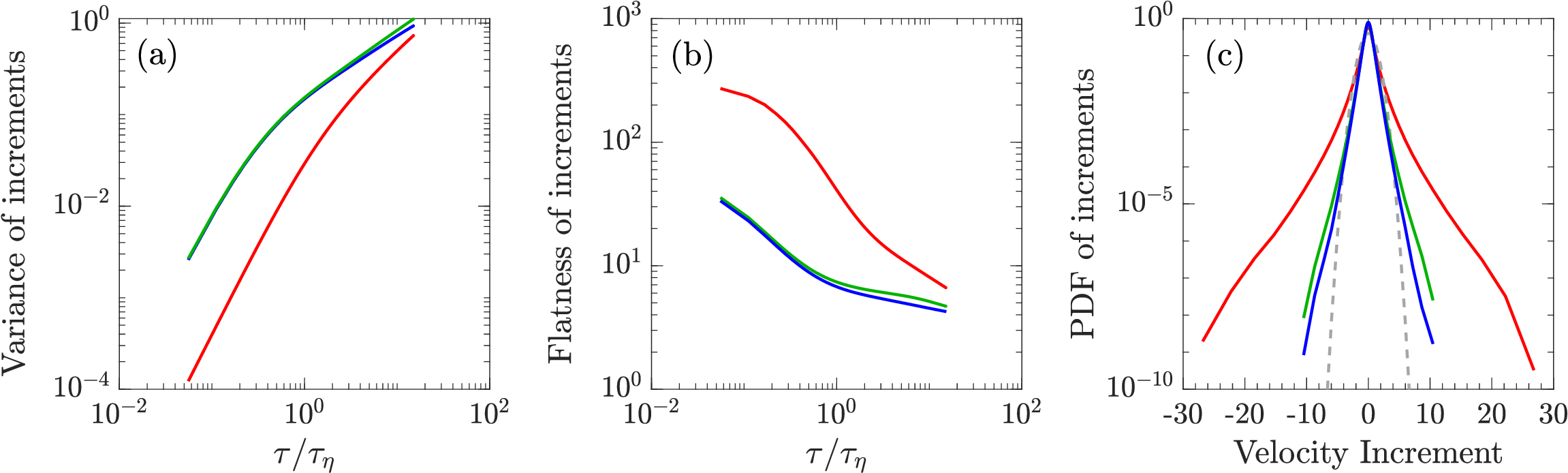}
\caption{\small (a) Variance, (b) Flatness factor and (c) PDF 
	(at $\tau/\ktime\approx 4$)
	of the Lagrangian velocity increment (red), and its
	convective (green) and local (blue) contributions at $\re\sim1300$.}
\label{fig:flatness_pdf}
\end{figure}

In the discussion of Figs.~\ref{fig:cpdf_re390}-\ref{fig:cpdf_re1300}
we have focused on samples of $\uv_C$ and $\uv_L$ in the order
of several standard deviations, which
provide dominant contributions to lower order moments
including variances and covariances.
To understand how the mutual cancellation between $\uv_C$ and $\uv_L$
relate to the evolution of intermittency associated with larger
fluctuations, we show in
Fig.~\ref{fig:flatness_pdf} the dependence on $\tau$ for
 $\re\sim$ 1300 data in terms of
the variance and flatness factor of
$\Delta_\tau \uu^+$, $\uv_C$ and $\uv_L$, as well
as the PDFs of all three increments at a selected time lag.
The flatness factor is high at small $\tau$ and decreases with $\tau$ towards the Gaussian limit (3.0) for all three increments, showing that extreme samples primarily occur at short time lags, in agreement with the small $\tau$ limit of $\Delta_\tau \uu^+\approx (\ua_C+\ua_L)\tau$. We also see that,
in accordance with prior findings for the Lagrangian
acceleration $\ua$ and its contributions $\ua_C$ and $\ua_L$
\citep{TVY.2001}
the total increment $\Delta_\tau\uu^+$ is more intermittent than both $\uv_C$ and $\uv_L$,
although, the contrast becomes weaker as $\tau$ increases.
The statistics of $\uv_C$ and $\uv_L$ are very similar,
but $\uv_C$ is slightly more intermittent than $\uv_L$.
It is thus not surprising that, when forming $\Delta_\tau\uu^+$,
large values of $\uv_L$ are prone to be canceled out
by accompanying similarly large or even larger values of $\uv_C$,
while the converse is less probable. The overall result is that 
some large samples of $\uv_C$ and $\uv_L$ may survive the mutual cancellation,
and in conjunction with the possibility of a positive
alignment angle (as seen in Fig.~\ref{fig:corcof}), 
contribute to forming extreme values of $\Delta_\tau \uu^+$ despite the overall statistical mutual cancellation effect of random-sweeping.
The velocity increment PDFs in frame (c) are taken 
at $\tau/\ktime\approx 4$, which is close to the time lag
where approximate Kolmogorov scaling for the
velocity structure function starts.
At this time lag, the PDFs of $\uv_C$ and $\uv_L$
are actually not far from Gaussian, but it is clear
that the PDF of 
$\Delta_\tau \uu^+$ shows much wider tails,
with samples detected up to about 25 standard deviations.

A primary finding in this subsection is that,
despite the apparent success of Tennekes' random-sweeping
hypothesis, mutual cancellation between convective
and local contributions to the Lagrangian velocity
increment is inherently incomplete, with the
incompleteness actually contributing to strong
intermittency. In the next subsection, we examine  the 
convective increment in more detail, with a focus on the effect of
the particle displacement, 
as a random parameter in how the convective increment
is calculated.


\subsection{The role of particle displacements in Lagrangian structure functions}

\def\ul {\boldsymbol{\ell}}

Let $\ul(\tau)$ be the fluid particle displacement vector,
and $\ell$ be its length, measured over a time increment $\tau$. 
In isotropic turbulence 
$\langle \ell^2(\tau)\rangle =\langle \ell_i\ell_i\rangle=3\sigma_1^2(\tau)$,
where each Cartesian component of $\ul$ has
the same variance $\sigma_1^2(\tau)$ and being the integral
of a Gaussian-distributed velocity component is thus itself close to Gaussian. As a result,
$\ell^2/\sigma_1^2$ has a chi-squared distribution of order~3, while its square root $\ell/\sigma_1$ follows a so-called chi distribution,
with its PDF being
\be
f_{\ell/\sigma_1}(x) = \frac{1}{\sqrt{2}\Gamma(3/2)} x^2\exp(-x^2/2),
\ee
where $\Gamma(\cdot)$ is the Gamma function, and
$\Gamma(3/2)=\sqrt{\pi}/2$. We are interested in the probability
that $\ell/\sigma_1$ exceeds a certain threshold (say $x$), which
is given by a complementary
cumulative distribution function (CCDF) of the form
\be\label{eq:CDF1}
P(\ell/\sigma_1>x) = 1-F_{\ell/\sigma_1}(x) = \frac{\Gamma^{*}(3/2,{x^2/2})}{\Gamma(3/2)},
\ee
where $F_{\ell/\sigma_1}$ is the cumulative distribution function (CDF) of $\ell/\sigma_1$, and 
${\Gamma^{*}(3/2,x^2/2}) = \int_{x^2/2}^{\infty}t^{1/2}\exp(-t)~dt$
is the upper incomplete gamma function (of order 3/2).
The CCDF of the
Kolmogorov-scaled displacement $\ell/\eta$ can also be obtained
by using the relation 
$P(\ell/\eta > x) = P(\ell/{\sigma_1} > ax) = 1 - F_{\ell/\sigma_1}(ax)$
where $a=\eta/\sigma_1$.

It is readily understood that at small times
(the ballistic range)  
$\sigma_1^2\approx u'^2\tau^2$ where $u'$ is the RMS velocity,
while at large times (in the diffusion range)
$\sigma_1^2\approx 2u'^2 T_L \tau$.
In the ballistic range we can also compare $\sigma_1$ to $\eta$ by writing
\be
\frac{\sigma_1}{\eta} 
\approx
\frac{u'\tau}{\eta}
=\frac{u'}{u_\eta}\frac{\tau}{\ktime}
\approx \left(\frac{1}{15}\right)^{1/4}\sqrt{\re}
\left(\frac{\tau}{\ktime}\right),
\ee
where the only assumption required is the standard local isotropy relation $\epsav=15\nu (u'/\lambda)^2$.
The appearance of the $\sqrt{\re}$ factor here implies that,
especially at high Reynolds number, the distance traveled
can quickly become large compared to the Kolmogorov scale
--- venturing even into the inertial range 
--- at relatively small $\tau/\ktime$.

\begin{figure}[h]
\centering
{\includegraphics[width=0.9\textwidth]{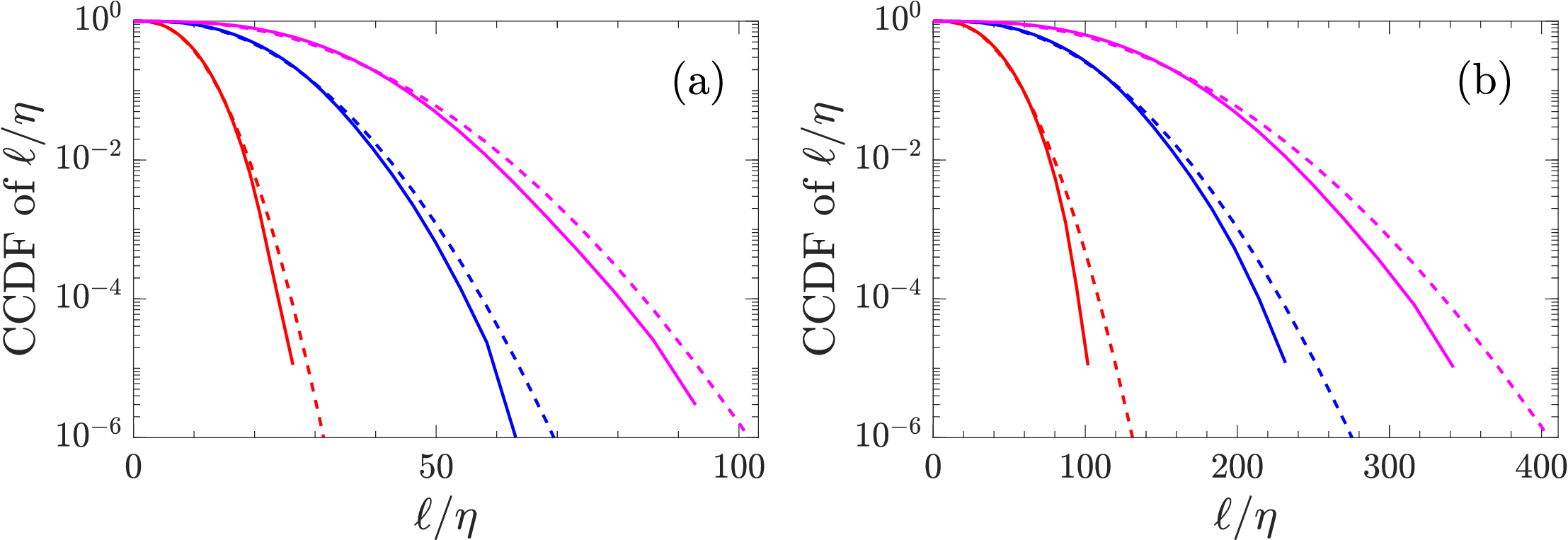}}
\caption{CCDFs of Kolmogorov-scaled particle displacement $\ell/\eta$ from DNS data at $\re\sim 140$ (red), $650$ (blue), $1300$ (magenta), at $\tau/\ktime\approx 1$ (frame a) and 4 (frame b). Dashed lines represent a theoretical estimate based on Eq.~\eqref{eq:CDF1}. 
}
\label{fig:cdfpdsp}
\end{figure}

Figure~\ref{fig:cdfpdsp} shows the CCDFs of $\ell/\eta$ at two values of time lags for three Reynolds numbers.
It can be seen in Fig.~\ref{fig:cdfpdsp}(a), that within just 1 $\ktime$ over 50\% of the particles
have moved by roughly 9 $\eta$ if $\re$ is 140, increasing to almost 30 $\eta$
if $\re$ is 1300. The observation that a large number of particles sample large values of $\ell$ within a short time lag is consistent with the results of~\cite{homann2009bridging}. In addition, in the $\re\sim$ 1300 simulation
within 4 $\ktime$, 80\% of the particles
have traveled at least 110 $ \eta$, which is long enough
to meet the requirements
for inertial range in many aspects of Eulerian statistics.
At later times, the shift towards larger displacement values
slows down somewhat, as a result of $\sigma_1^2$ becoming
proportional to $\tau$ in the diffusive range as opposed to $\tau^2$ in the ballistic range. Nevertheless,
at sufficiently large $\tau$ many particles will likely
have traveled by distances comparable to the integral length scale of the flow. 
At large $\ell/\eta$, the solid and dashed curves do not fully
agree, which can be related to the fact that 
the independence between different coordinate components
assumed in the chi distribution is not guaranteed in actual
DNS results where sampling is finite. The trend of 

{
\begin{figure}[h]
\centering
{\includegraphics[width=0.9\textwidth]{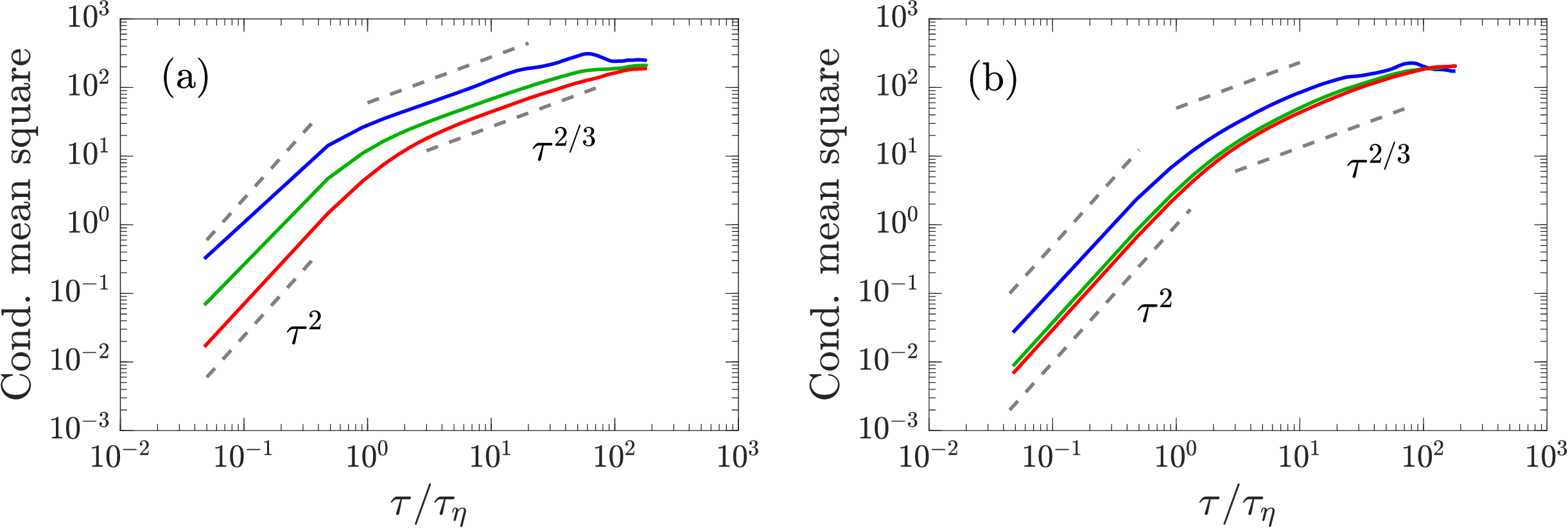}}
\caption{\small Conditional mean-square of (a) convective increment 
and (b) total Lagrangian increment of velocity fluctuations, given 
displacement $\ell/\langle\ell\rangle$ normalized by instantaneous 
mean values at close to 1/2 (red), 1 (green), 2 (blue), at $\re\sim$ 390. All axes are normalized by Kolmogorov variables.
Short dashed lines represent power laws of exponents 2 and 2/3.
}
\label{fig:condmsq_re390}
\end{figure}
}

To clarify the effect of particle displacements on the statistics
of the velocity increments, we have computed the mean squares 
of $\uv_C$, $\uv_L$  and $\Delta_\tau\uu^+$ 
conditioned on the displacement $\ell$ as a function of $\tau$.
Figure~\ref{fig:condmsq_re390} shows the conditional mean squares
given $\ell/\ellav$ in the neighborhoods of 0.5, 1, and 2,
with both axes scaled by Kolmogorov variables, using data from the $\re\sim$ 390 simulation,  for
$\uv_C$ and $\Delta_\tau\uu^+$ in frames (a) and (b) respectively. In frame (a), the contrasts between lines of different colors
indicate that the convective increment increases strongly
with $\ell/\ellav$ at most values of $\tau$. 
For small $\ell/\ellav$, the conditional mean squares of $\uv_C$
show regimes of ballistic increase (slope 2 on logarithmic
scales) at small $\tau$, inertial range behavior (slope 
close to 2/3 associated with the scaling of Eulerian structure functions) at intermediate  $\tau$, and finally
slower increases at large $\tau$.
However, for larger $\ell$, the end of the ballistic regime,
the start of the inertial range, and weaker growth at later 
times, all occur at smaller values of $\tau/\ktime$ compared to 
the case of small $\ell$. This contrast is not surprising, since 
a larger value of $\ell$ as the conditioning variable implies samples are taken over positions that are further apart.
At the same time, with the skewness of the chi 
distribution for $\ell/\ellav$ being small~{($\sim0.49$)}, the conditional mean square given $\ell/\ellav\approx 1$ agrees with the unconditional result very
well (to within less than 1\%, hence not shown here).

Frame (b) of Fig.~\ref{fig:condmsq_re390} shows the
conditional mean square of the total increment
$\Delta_\tau\uu^+$ follows similar trends, but also 
displays some differences. Specifically, ballistic behavior in 
the blue line is better captured, which is consistent with
the effect of large particle displacements
in the convective increment being partly canceled by
the local increment that also contributes to
$\Delta_\tau\uu^+$. The sensitivity of
the statistics of $\Delta_\tau\uu^+$ to $\ell/\ellav$, inferred from the spacings among curves of different colors, is significantly 
weaker than that seen in the statistics of $\uv_C$ (in frame 
(a)). It is also clear that, regardless of the conditioning
$\ell/\ellav$, as $\tau/\ktime$ approaches inertial-range
values, the scaling of the total increment is
more ambiguous, since the spatial-temporal Lagrangian velocity increment scales differently from either the spatial or temporal increments.

\begin{figure}[h]
\centering
{\includegraphics[width=0.9\textwidth]{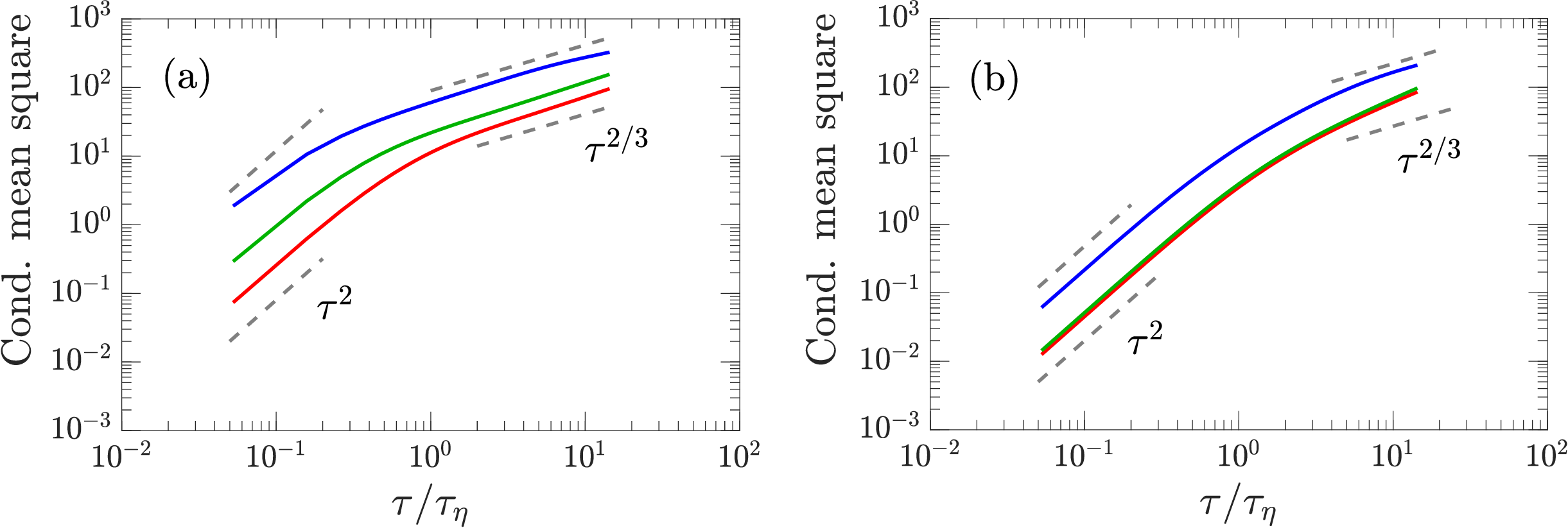}}
\caption{\small Same as Fig.~\ref{fig:condmsq_re390}, but showing data at $\re\sim$ 1300.
}
\label{fig:condmsq_re1300}
\end{figure}

To assess Reynolds number effects on the conditional statistics,
in Fig.~\ref{fig:condmsq_re1300} we show results
at $\re\sim$ 1300, where stronger intermittency is accompanied by
a time scale ratio $T_L/\ktime$ more than three times larger
(per Table~I) than that at $\re\sim$ 390.
In frame (a), even at small $\tau$,
the wider range of scales can cause
the mean square of $\uv_C$ given $\ell/\ellav\approx 2$ (or higher)
to depart quickly from ballistic
behavior, since such values of $\ell$ can quickly
become much larger than $\eta$. However, at intermediate time lags,
$\tau^{2/3}$ scaling for $\re\sim$ 1300 is noticeably
clearer than at $\re\sim$ 390. In contrast, the conditional
mean square in frame (b) closely resembles that for $\re\sim$ 390
in frame (b) of Fig.~\ref{fig:condmsq_re390} earlier,
except that the $\re\sim$ 1300 simulation only reached 15~$\ktime$ in time.
The relative lack of a convincing power law in the 
conditional mean squares of
$\Delta_\tau\uu$ (versus that for $\uv_C$)
is also consistent with uncertainties in the application
of traditional K41 similarity to the second-order 
Lagrangian structure function addressed in Sec.~III.

The results in Figs.~\ref{fig:condmsq_re390} and~\ref{fig:condmsq_re1300}, as discussed above, raise the question of whether the unconditional mean square may also
display any scaling ranges of the type observed for the conditional quantities.
Since the $\tau^{2/3}$ scalings in the conditional
statistics above have been obtained only
from fluid particles that
have traveled by a certain distance in the inertial range
over a given time interval, their relevance
to the global statistics over all the particles
depends on the probability of $\ell$ being in the
inertial range (and the extent of such).
This probability can be estimated by subtracting
between values of the CCDF of $\ell/\eta$ corresponding to
values of $r/\eta$ that mark the start and end of the inertial
range. For $\re\sim$ 1300, based on data on the third-order
Eulerian structure functions~\cite{ISY.2017}
we can take the inertial range available as
$50\leq r/\eta\leq 400$.
The data in Fig.~\ref{fig:cdfpdsp} gives
the proportion of fluid particles in this range
as 5 \% at $\tau/\ktime\approx 1$ 
and 94 \% at $\tau/\ktime\approx 4$.
The high proportion in the latter case strongly suggests that the
unconditional mean square of $\uv_C$ should behave
similarly as their conditional counterparts
in Fig.~\ref{fig:condmsq_re1300}(a). 

\begin{figure}[ht]
    \centering
    \includegraphics[trim={0in 0in 0in 0in},clip,width=0.9\textwidth]{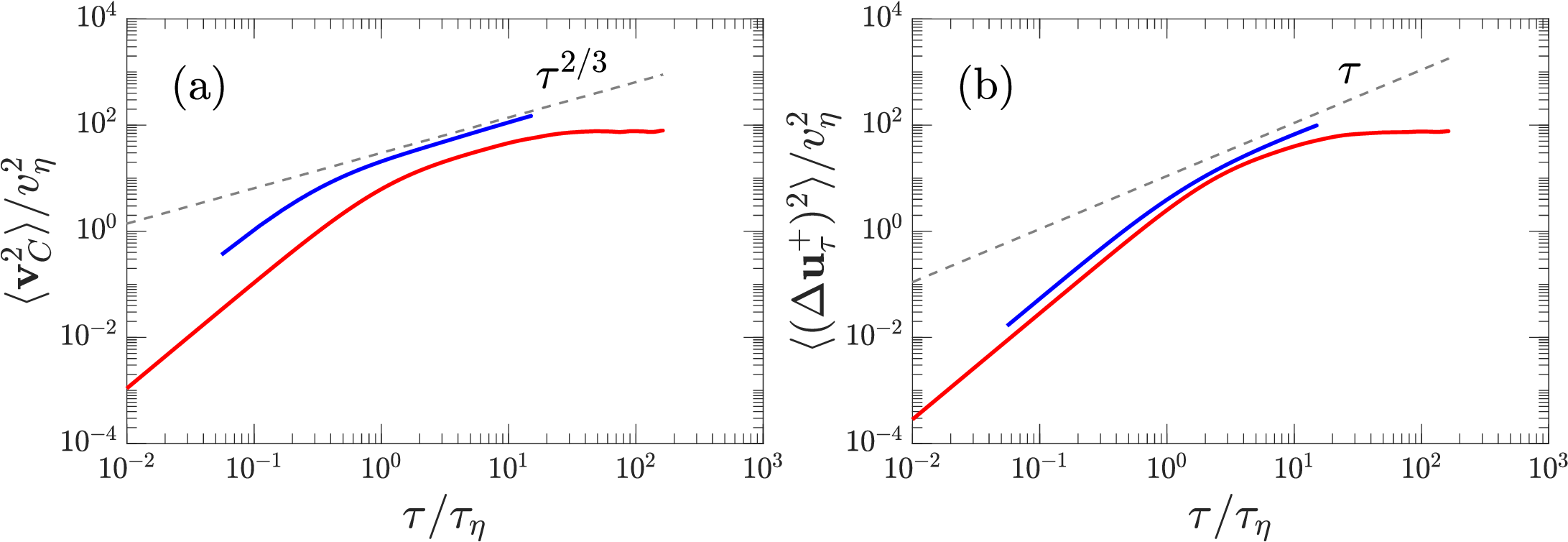}
    \caption{Evolution of $\big\langle(\Delta_\tau \uu^+)^2\big\rangle$ and $\big\langle \uv_C^2\big\rangle$ in time at $\re\sim 140$ (red) and $1300$ (blue), scaled by Kolmogorov variables. Dashed lines of slope $1$ and $2/3$ are shown for comparison.}
    \label{fig:scaling}
\end{figure}

The connection between the 
$r^{2/3}$ scaling for the mean-squared spatial velocity increment $\Delta_r\uu$ in the Eulerian inertial range,
and the inertial range scaling for the mean-squared 
spatial increment $\uv_C$ merits some additional
discussion. While for sufficiently small $\tau$, a Taylor-series
expansion indicates that $\ell\propto\tau$ (statistically), with the 
proportionality factor being the RMS velocity, if $\tau$ is several
$\ktime$ long, the 
Taylor series argument may require justification. However, at high $\re$, a time lag of a few $\ktime$ can perhaps still be considered to be small compared to $T_L$. Thus, if the velocity 
autocorrelation remains high at the time lags involved, the 
Taylor series approximation may sufficiently hold. It may also be argued that since power laws are defined by local slopes, the Taylor 
series argument remains applicable if the question is only whether a portion of the curve showing 
$\langle\uv_C^2\rangle$, where the $\ell$ involved corresponds to separations in the Eulerian inertial range, displays the logarithmic slope of 2/3. In other words,
since the convective velocity increment $\uv_C$ can be interpreted as 
a velocity increment over the spatial separation $\ell(\tau)$, its variance 
should scale as $\tau^{2/3}$ provided $\ell$ lies in the Eulerian inertial range. The Reynolds number trends observed in 
Fig.~\ref{fig:cdfpdsp} also suggest that such a $\tau^{2/3}$ scaling region would emerge earlier and last longer as $\re$ increases.

In Fig.~\ref{fig:scaling} we show the evolution
of the unconditional mean-squared
(a) convective and (b) total velocity increments
scaled by the Kolmogorov variables.
For $\re\sim$ 1300 the convective increments  indeed show
$\tau^{2/3}$ behavior --- starting from $\tau/\tau_\eta$ just 
slightly beyond~1.0 --- for the reasons given above, based on both the Taylor series approximation as well as the conditional mean squares.
For $\re\sim$ 140 this trend starts later because of $u'/u_\eta$
being smaller at lower $\re$, so that it takes a larger
number of time lags (in units of $
\ktime$) for most particles to have traveled
by a distance in the inertial range. This is also consistent with our observation in Fig.~\ref{fig:cdfpdsp} that particle displacements reach separations in the inertial range at progressively smaller $\tau$ as the $\re$ increases. The $\re\sim$ 140 simulation was also sufficiently long
(at least in units of $\ktime$) for the curve to eventually
level out when most of the particles are so far
apart that $\uu(\ux^+(t+\tau),t+\tau)$ and 
$\uu(\ux^+(t),t+\tau)$ become fully independent.
While the $\re\sim$ 1300 simulation only reached about 15 $\ktime$,
the same behavior can be expected at late times, implying that
the $\tau^{2/3}$ scaling only lasts for a modest length of time.
On the other hand, in frame (b), the scaling that is observed, 
although only approximately, in the range for several $\ktime$ is $\tau^1$, which corresponds to the classical K41 result for the second-order Lagrangian structure function.
As noted earlier, the dissimilarity between the scaling
of the convective and total increments is an indication
of the strength of the (partial) mutual cancellation that exists
between convective and local contributions.
The choice of logarithmic versus linear scales is the main reason
why sensitivity to Reynolds number in Fig.~\ref{fig:scaling}
looks much weaker than that seen in Fig.~\ref{fig:strfn}. We also note in passing that in the ballistic
range, all the curves in Fig.~\ref{fig:scaling}
scale as $\tau^2$. The heights at a given small $\tau/\ktime$
are proportional to the respective acceleration variances,
which increase with Reynolds number. The Reynolds number
dependence of $\langle\uv_C^2\rangle$ is also stronger
than that of $\langle(\Delta_\tau (\uu^+)^2\rangle$,
as reported previously in~\cite{TVY.2001}.

The analyses provided in this section show that particle
displacement plays a significant role in the scaling of the Lagrangian structure function, which is made clear
when viewed in a spatial-temporal framework.
However, the strong partial cancellation between the
convective and local increments leads to the total
Lagrangian velocity increment behaving differently,
including showing a different approximate power law
at intermediate times.

\section{Conclusions}

A well-known prediction from the application of classical Kolmogorov theory in the Lagrangian
frame is that the second-order velocity structure function should scale linearly 
(Eq.~\eqref{eq:C0}) over
intermediate time lags ($\tau$), at sufficiently high Reynolds number. 
However, unambiguous evidence 
showing the associated scaling constant ($C_0$)
approaching an asympotic value at high Reynolds
number has been difficult to find. In particular, in direct numerical simulations (DNS), the normalized structure function often possesses a local peak (rather than a plateau)
which forms after just 5-10 Kolmogorov time scales ($\ktime$) 
but is subsequently quickly truncated. Although one contributing factor  is that the range of time scales widens more slowly with
Reynolds number than the range of length scales, there is
a need for improved understanding, including why the
scaling range forms so quickly and only lasts for
a short amount of time.
In this paper, we have used DNS data in  forced isotropic turbulence 
at Taylor-scale Reynolds numbers ($\re$) 140 to 1300,
with connections to
the statistics of fluid particle acceleration
and the effects of particle displacement ($\ell(\tau)$) 
from a spatial-temporal perspective.

To assess statistical aspects of uncertainties related to
finite simulation time span, we have considered
the mean-squared velocity increment as a double integral 
of the acceleration autocorrelation, whose properties,
including a quick zero-crossing followed by a slow return
to zero, are well known. This analysis shows that the peak
value ($C_0^*$, in Eq.~\eqref{eq:C0_gamma}) of the 
Kolmogorov-scaled second-order structure function
has a strong connection to the Kolmogorov-scaled acceleration
variance ($a_0$), which increases with Reynolds number due to intermittency.
It also appears that, in practice, for the type of forced turbulence considered,
a simulation spanning $10~\ktime$ can provide reasonable accuracy for 
key results including $C_0^*$ in this work.
The current data does not clearly distinguish between the viability
of power laws (with no asymptotic constancy) and interpolation
formulas that assume asymptotic constancy at high Reynolds number.
However, if asymptotic constancy does occur it will 
require Reynolds
numbers not attainable in the foreseeable future, and the
value concerned is likely higher than prior estimates in the literature.

As written in Eq.~\eqref{eq:stdecomp}, 
the Lagrangian velocity increment
$\Delta_\tau\uu^+$ is inherently a spatial-temporal
quantity with convective ($\uv_C$) and local ($\uv_L$) contributions, that are subject to strong yet incomplete mutual
cancellation, in agreement with the so-called random-sweeping hypothesis. The partial cancellation results in 
$\Delta_\tau\uu^+$ having a smaller variance
but higher intermittency than both $\uv_C$ and $\uv_L$, especially at higher
Reynolds numbers, although the effect weakens at later times.
A more intriguing aspect is the role of the particle displacement,
which is well described by a chi distribution
of order three. Since $\ell(\tau)$ (i.e., the change of $\ell$ over 
any short time interval) scales with the RMS velocity $u'$, changes of $\ell$ much larger than $\eta$ can occur even for $\tau$ only at a few $\ktime$. 
Statistics of convective increments conditioned upon $\ell/\eta$
help explain the emergence of a convincing $\tau^{2/3}$ 
scaling in $\langle\uv_C^2\rangle$, in connection with the $r^{2/3}$
scaling for Eulerian velocity structure functions in the inertial range.
For instance, in our data at $\re\sim$ 1300,
the proportion of samples of $\ell$ within an estimated inertial range
of spatial separations between 50 and 400 Kolmogorov length scales
rises from 5\% at $\tau/\ktime\approx 1$ to as high as 94\% at
$\tau/\ktime\approx 4$. This high percentage implies that the mean-square
of $\uv_C$ behaves similarly to its own conditional value
given $\ell$ in the appropriate scaling range.
Finally, although the incomplete cancellation implies that $\langle(\Delta_\tau \uu)^2\rangle$ does not follow
the same power law, the occurrence of large displacements
within short time intervals still helps explain why the structure
function starts to exhibit inertial-like scaling as early
as a few $\ktime$, and that the approximate scaling is also
quickly washed out as particle displacements begin to 
exceed inertial scales in space.

In summary, we believe this work provides new insights into the
observed behavior and often limited or elusive inertial range
scaling for the second-order Lagrangian structure function,
including whether the Lagrangian Kolmogorov constant
will approach an asymptotic constant value in the limit of
very large Reynolds number. A spatial-temporal
perspective shows that both the limited range of time scales (compared to length scales) and the time-dependent particle displacements play significant roles. 
A related open question is how at least some details of the Lagrangian dynamics would differ when particle inertia is taken into account. This question may also be of interest when interpreting experimental data
based on following the trajectories of tracer particles, which may not exactly move with the flow.

\bibliography{apssamp}
\begin{acknowledgments}
This work was partially supported by the National Science Foundation's
Computational and Data-Enabled Science and Engineering Program, Grant 1953186.
The authors acknowledge the Texas Advanced Computing Center (TACC) at The University of Texas at Austin for providing computational resources that have contributed to the research results reported within this paper. URL: http://www.tacc.utexas.edu. We also thank Professors Stephen B. Pope, Brian Sawford and Michael Wilczek for helpful discussions and valuable comments.
\end{acknowledgments}

\end{document}